\documentclass[]{aastex631}

\usepackage{bm}
\usepackage{rotating}

\shorttitle{Large amplitude bidirectional anisotropy}
\shortauthors{Munakata et al.}
\begin{document}

%\begin{potrait}

\title{Large amplitude bidirectional anisotropy of cosmic-ray intensity observed with world-wide networks of ground-based neutron monitors and muon detectors in November, 2021}

\author[0000-0002-2131-4100]{K. Munakata}\affiliation{Department of Physics, Shinshu University, Matsumoto, Nagano 390-8621, Japan}
\author[0000-0002-3948-3666]{M. Kozai}\affiliation{Polar Environment Data Science Center, Joint Support-Center for Data Science Research, Research Organization of Information and Systems, Tachikawa, Tokyo 190-0014, Japan}
\author[0000-0002-4913-8225]{C. Kato}\affiliation{Department of Physics, Shinshu University, Matsumoto, Nagano 390-8621, Japan}
\author[0000-0002-0890-0607]{Y. Hayashi}\affiliation{Department of Physics, Shinshu University, Matsumoto, Nagano 390-8621, Japan}
\author[0000-0001-9400-1765]{R. Kataoka}\affiliation{National Institute of Polar Research, Tachikawa, Tokyo 190-8518, Japan}
\affiliation{Department of Polar Science, School of Multidisciplinary Sciences,The Graduate University for Advanced Studies, SOKENDAI, Tachikawa, Tokyo 190-8518, Japan}
\author[0000-0002-6105-9562]{A. Kadokura}\affiliation{Polar Environment Data Science Center, Joint Support-Center for Data Science Research, Research Organization of Information and Systems, Tachikawa, Tokyo 190-0014, Japan}
\affiliation{National Institute of Polar Research, Tachikawa, Tokyo 190-8518, Japan}
\affiliation{Department of Polar Science, School of Multidisciplinary Sciences,The Graduate University for Advanced Studies, SOKENDAI, Tachikawa, Tokyo 190-8518, Japan}
\author[0000-0002-2982-1887]{M. Tokumaru}\affiliation{Institute for Space-Earth Environmental Research, Nagoya University, Nagoya, Aichi 464-8601, Japan}
\author[0000-0002-0417-6877]{R. R. S. Mendonça}\affiliation{National Institute for Space Research (INPE), 12227-010 São José dos Campos, Brazil}
\author[0000-0002-8351-6779]{E. Echer}\affiliation{National Institute for Space Research (INPE), 12227-010 São José dos Campos, Brazil}
\author[0000-0002-4361-6492]{A. Dal Lago}\affiliation{National Institute for Space Research (INPE), 12227-010 São José dos Campos, Brazil}
\author[0000-0002-9737-9429]{M. Rockenbach}\affiliation{National Institute for Space Research (INPE), 12227-010 São José dos Campos, Brazil}
\author[0000-0002-7720-6491]{N. J. Schuch}\affiliation{Southern Space Coordination, National Institute for Space Research, P.O. Box 5021 - 97110-970 - Santa Maria, RS - Brazil}
\author[0000-0003-2931-8488]{J. V. Bageston}\affiliation{Southern Space Coordination, National Institute for Space Research, P.O. Box 5021 - 97110-970 - Santa Maria, RS - Brazil}
\author[0000-0003-1485-9564]{C. R. Braga}\affiliation{George Mason University, 4400 University Drive, Fairfax, Virginia 22030, USA}
\author[0000-0002-4486-237X]{H. K. Al Jassar}\affiliation{Physics Department, Kuwait University, P.O. Box 5969 Safat, 13060, Kuwait}
\author[0000-0002-2090-351X]{M. M. Sharma}\affiliation{Physics Department, Kuwait University, P.O. Box 5969 Safat, 13060, Kuwait}
\author[0000-0001-7463-8267]{M. L. Duldig}\affiliation{School of Natural Sciences, University of Tasmania, Hobart, Tasmania 7001, Australia}
\author[0000-0002-4698-1672]{J. E. Humble}\affiliation{School of Natural Sciences, University of Tasmania, Hobart, Tasmania 7001, Australia}
\author[0000-0002-7037-322X]{I. Sabbah}\affiliation{Department of Natural Sciences, College of Health Sciences, Public Authority of Applied Education and Training, Kuwait City 72853, Kuwait}
\author[0000-0001-7929-810X]{P. Evenson}\affiliation{Bartol Research Institute and Department of Physics and Astronomy, University of Delaware, 217 Sharp Laboratory, Newark, DE 19716, USA}
\author[0000-0003-4865-6968]{P.-S. Mangeard}\affiliation{Bartol Research Institute and Department of Physics and Astronomy, University of Delaware, 217 Sharp Laboratory, Newark, DE 19716, USA}
\author[0000-0001-6584-9054]{T. Kuwabara}\affiliation{Bartol Research Institute and Department of Physics and Astronomy, University of Delaware, 217 Sharp Laboratory, Newark, DE 19716, USA}
\author[0000-0003-3414-9666]{D. Ruffolo}\affiliation{Department of Physics, Faculty of Science, Mahidol University, Bangkok 10400, Thailand}
\author[0000-0001-7771-4341]{A. S\'aiz}\affiliation{Department of Physics, Faculty of Science, Mahidol University, Bangkok 10400, Thailand}
\author[0000-0002-3776-072X]{W. Mitthumsiri}\affiliation{Department of Physics, Faculty of Science, Mahidol University, Bangkok 10400, Thailand}
\author[0000-0002-1664-5845]{W. Nuntiyakul}\affiliation{Department of Physics and Materials Science, Faculty of Science, Chiang Mai University, Chiang Mai 50200, Thailand}
\author[0000-0002-3715-0358]{J. K\'ota}\affiliation{Lunar and Planetary Laboratory, University of Arizona, Tucson, AZ 85721, USA}

%\author[0000-0002-0786-7307]{Greg J. Schwarz}
%\affiliation{American Astronomical Society \\
%1667 K Street NW, Suite 800 \\

%\author{August Muench}
%\affiliation{American Astronomical Society \\
%1667 K Street NW, Suite 800 \\

%\collaboration{6}{(AAS Journals Data Editors)}

%\author{Butler Burton}
%\affiliation{Leiden University}
%\affiliation{AAS Journals Associate Editor-in-Chief}

%\altaffiliation{AASTeX v6+ programmer}

%\affiliation{AAS Director of Publishing}
%\affiliation{American Astronomical Society \\
%1667 K Street NW, Suite 800 \\

%\author{Magaret Donnelly}
%\affiliation{IOP Publishing, Washington, DC 20005}



\begin{abstract}

We analyze the cosmic-ray variations during a significant Forbush decrease observed with world-wide networks of ground-based neutron monitors and muon detectors during November 3-5, 2021. Utilizing the difference between primary cosmic-ray rigidities monitored by neutron monitors and muon detectors, we deduce the rigidity spectra of the cosmic-ray density (or omnidirectional intensity) and the first- and second-order anisotropies separately, for each hour of data. A clear two-step decrease is seen in the cosmic-ray density with the first $\sim2\%$ decrease after the interplanetary shock arrival followed by the second $\sim5\%$ decrease inside the magnetic flux rope (MFR) at 15 GV. Most strikingly, a large bidirectional streaming along the magnetic field is observed in the MFR with a peak amplitude of $\sim5\%$ at 15 GV which is comparable to the total density decrease inside the MFR. The bidirectional streaming could be explained by adiabatic deceleration and/or focusing in the expanding MFR, which have stronger effects for pitch angles near 90$^\circ$, or by selective entry of GCRs along a leg of the MFR. The peak anisotropy and density depression in the flux rope both decrease with increasing rigidity. The spectra vary dynamically indicating that the temporal variations of density and anisotropy appear different in neutron monitor and muon detector data.
%This is the first attempt to quantitatively deduce dynamic variations of rigidity spectra of the cosmic-ray anisotropy including the second-order anisotropy during a single event by analyzing the neutron monitor and muon detector data together.
  
\end{abstract}


\section{Introduction} \label{sec:intro}

The galactic cosmic-ray (GCR) intensity observed at Earth dynamically changes in association with the arrival of the solar and interplanetary disturbances, such as the interplanetary coronal mass ejections (ICMEs) with or without an interplanetary (IP) shock and the corotating interaction regions (CIRs). The Forbush decrease (FD) is the most well-known phenomenon representing the dynamical change of GCR intensity. Since the GCR variation observed at a point in space consists of two different superposed components, the variation of the GCR density (omnidirectional intensity) and the variation due to the anisotropy, the multidirectional observations using a global detector network are necessary to study these components separately and accurately. For this purpose, the world-wide network observations with the ground-based detectors have been employed.\par

Recent examples of analyses of the neutron monitor network data are given by \citet{Belov18} and \citet{Abunin20}. The neutron monitors (NMs), which detect secondary neutrons produced by GCRs interacting with atmospheric nuclei, have a maximum response to primary GCRs with median rigidities between $\sim10$ GV and $\sim30$ GV. We define the median rigidity as the rigidity of primary GCRs below which the integrated rigidity response is a half of the total integrated response. The Global Muon Detector Network (GMDN), on the other hand, was established in 2006 with four multidirectional surface muon detectors at Nagoya in Japan, Hobart in Australia, Kuwait City in Kuwait and S\~ao Martinho in Brazil. Because a higher primary energy is needed to produce muons with sufficient Lorentz factor and relativistic time dilation to reach ground level before decaying, muon detectors (MDs) have a response to primary GCRs with higher median rigidities, between $\sim50$ GV and $\sim100$ GV. While the NM is an omnidirectional detector and it observes on average the vertically incident GCRs to the detector, a single MD can be multidirectional because the incident direction of muons better preserves the incident direction of primary GCRs at the top of the atmosphere. Based on the diffusive transport picture of GCRs in which the first-order anisotropy (or diurnal anisotropy) is directly related to the spatial gradient of GCR density, the diurnal anisotropy observed by GMDN has been used to deduce the ICME geometry and orientation from the density gradient \citep{Kuwa09}. A summary of studies using the GMDN data can be found in \citet{Marlos14}.\par

While most of the preceding studies using the NM network and GMDN separately have been limited to analyze the variations of the GCR density and the first-order anisotropy\citep{Belov18,Tortermpun18,Kihara20}, the second-order anisotropy also has a significant contribution to the GCR variation in some events. Bidirectional streaming (BDS) is often observed in the satellite measurements of low energy electrons and ions indicating particles trapped in the magnetic flux rope (MFR) in which the field line is anchored on Sun at both ends. By analyzing the NM network data together with the satellite data, \citet{Richardson00} found significant second order anisotropy due to the BDS of GCRs in magnetic clouds associated with ICMEs in 1982.\par

\citet{Ruffolo06} reported BDS of GCRs during a Forbush decrease and also during a ground level enhancement (GLE) of relativistic protons observed by a network of polar NMs on October 22, 1989.  The GLE is due to protons accelerated by a solar storm, as indicated by measurement of a soft rigidity spectrum, and injected along the interplanetary magnetic field (IMF) which is connected to Earth \citep{Danilova99}. They found two intensity peaks in the time profile of NM data and a strong second-order anisotropy during the second peak. Based on their numerical model of the particle transport along the IMF, and the BDS of GCRs, they concluded that the observations are consistent with an IMF configuration in which the field line is anchored at both ends to the Sun. \par

%\citet{Ruffolo06}, on the other hand, analyzed a ground level enhancement (GLE) of relativistic protons observed by a network of polar NMs in October 1989. These protons are accelerated during a solar flare and injected along the interplanetary magnetic field (IMF) which is connected to Earth \citep{Danilova99}. They found two intensity peaks in the time profile of NM data and a strong second-order anisotropy during the second peak. Interestingly, they also derived a soft rigidity spectrum of particles which is consistent with the solar energetic particles (SEPs) accelerated and injected by the flare and not GCRs. Based on their numerical model of the particle transport along the IMF, they concluded that the observation is consistent with an IMF configuration in which the field line is anchored at both ends on Sun. \par

In this paper, we report a large amplitude BDS observed in an ICME event during November 3-5, 2021 by analyzing the world wide network data of NMs and MDs. In particular, utilizing the difference between average cosmic-ray rigidities monitored by NMs and MDs, we derive the rigidity spectra of cosmic-ray density and anisotropy, each as a function of time. We describe the data and analysis in Sections \ref{subsec:data} and \ref{subsec:analyses}, respectively, and show the results in Section \ref{sec:result}. Finally, the summary and discussion are given in Section \ref{sec:discussion}\par

\section{Data and analyses} \label{sec:DandA}
To derive the rigidity spectra of the cosmic-ray density and anisotropy, we analyze hourly count rates recorded by worldwide networks of neutron monitors (NMs) and muon detectors (MDs) which respond to GCRs in different rigidity regions. In this section, we describe the data analyzed and the analysis method in the following subsections.\par

\subsection{Cosmic-ray data} \label{subsec:data}
In this paper, we analyze 90 hourly count rates recorded by 21 NMs and 69 directional channels of GMDN and Syowa MD (hereafter 69 MDs), which are available at websites \footnote{http://www01.nmdb.eu/} \footnote{http://www.thaispaceweather.com/} \footnote{https://cosray.shinshu-u.ac.jp/crest/DB/Documents/documents.php} \footnote{http://polaris.nipr.ac.jp/\~{}cosmicrays/}. The cosmic-ray data available from these websites are all corrected for the atmospheric pressure, while we added an additional correction to MD data for the atmospheric temperature effect by applying the method developed by \citet{Rafael16}. This method uses the mass weighted temperature calculated from the vertical profile of the atmospheric temperature provided by the Global Data Assimilation System (GDAS) of the National Center for Environmental Prediction available at the NOAA website\footnote{ftp:// ftp.arl.noaa.gov/archives/gdas1/}.\par

Table 1 lists characteristics of NMs and MDs including the geographical latitude, longitude and altitude of the detector's location ($\lambda_D$, $\phi_D$ and alt.), the number of available directional channels (ch-no.), the geomagnetic cut-off rigidity ($P_c$), the hourly count rate (cph), the count rate error ($\sigma$), the median rigidity of the detected primary GCRs ($P_m$) and the geographical asymptotic viewing direction outside the magnetosphere ($\lambda_{asymp}$ and $\phi_{asymp}$) for GCRs with $P_m$. We calculate $P_m$ as the rigidity of primary GCRs below which the integrated rigidity response is a half of the total integrated response. Our calculations of $P_m$, $\lambda_{asymp}$ and $\phi_{asymp}$ and the response functions used in our calculations are briefly described in Appendix A. We selected 21 NMs in Table 1 to maximize the sky coverage of viewing directions with least overlap.\par

We particularly include the data from PSNM (Princess Sirindhorn NM in Thailand) in operation at the world-highest $P_c$ monitoring high-energy GCRs. As seen in this table, $P_m$ ranges over 11.3-22.8 GV for data from 20 NMs, other than PSNM, while it ranges over 53.1-106.9 GV for MD data. The average $P_m$ weighted by the count rate is 14.9 GV for 20 NMs and 65.4 GV for MDs. Therefore, there is a factor of 4.4 difference between the average $P_m$ monitored by these 20 NMs and 69 directional channels of MDs. This  motivates the present work to derive the rigidity spectrum of the GCR variation by analyzing NM and MD data altogether. Since $P_m$ of PSNM is 34.6 GV, nearly half way between $P_m$ values monitored by 20 NMs and 69 MD directional channels, PSNM data may play an important role in evaluating analyses of the rigidity dependent modulation.\par

For another check of the analysis, we also include the data from a combined NM (SYOW) and MD (Syow-MD) which started simultaneous operations at the Antarctic Syowa station in 2018 \citep{Kato21}. Since orbits of vertically incident cosmic rays tend to be aligned with the field line in the polar geomagnetic field, the asymptotic viewing directions of a NM and the vertical channel of MD are similar within $\sim30^{\circ}$. This means that we can observe similar directions in space in different rigidity regions by using an NM and MD at a single location in the Antarctic. The data from SYOW and Syow-MD, therefore, can be used for further evaluation of the analysis in this paper.\par

\subsection{Analyses} \label{subsec:analyses}
Figures 1a-d show the solar wind parameters in an ICME recorded during November 3-5, 2021 when a large cosmic ray event was observed\footnote{https://omniweb.gsfc.nasa.gov/ow.html}. This ICME is related to an M1.7 X-ray flare that erupted near the central meridian of the Sun at 01:30 UT of November 2, and a halo CME observed by STEREO-A/COR 2 starting at 02:53 of the same day, which is one of multiple CME eruptions in November 1-2 when M and C-class X-ray flares were detected by {\it GOES} satellite, and the evidence of BiDirectional suprathermal Electron strahls (BDE) was also reported from satellite measurements for this event\footnote{https://izw1.caltech.edu/ACE/ASC/DATA/level3/icmetable2.htm}. A ground level enhancement (GLE) has been reported from the worldwide network of NMs on October 28 associated with an X-class flare\footnote{https://cosmicrays.oulu.fi/}, but no other enhancement is reported in October and November. Following an interplanetary shock (IP-shock) arrival indicated by a vertical orange line at around 20:00UT of 3 November, a clear magnetic flux rope (MFR) signature of a smooth rotation of the IMF orientation (black curve of panel b) is observed during a period between 12:00 November 4 and 06:00 November 5, delimited by a pair of vertical purple lines.\par

We analyze the percent deviation of the hourly count rate $I_{i,j}(t)$ recorded in the $j$-th directional channel of the $i$-th detector ($j=1$ for all NMs, $j=1-17$ for Nagoya and São Martinho da Serra MDs, $j=1-13$ for Hobart and Kuwait MDs and $j=1-9$ for Syowa MD) at universal time $t$ in hours from an average over a solar rotation period (CR2250) between 22 October and 17 November, 2021. We excluded two full days of October 28 and 29 from the analysis to avoid possible influence of a GLE in October 28. The bottom two panels of Figure 1 show a sample four $I_{i,j}(t)$ traces each among 21 NMs (Figure 1e) and 5 MDs (Figure 1f). It is seen in Figure 1e that mid-latitude NM data (ATHN, black curve) with high ${P_c}$ start to decrease after the shock arrival and reach a minimum of about -6\% in the MFR period delimited by a pair of purple vertical lines. A clear two-step decrease is seen in a polar NM (SYOW, purple curve), i.e., a first step in the magnetic sheath region after the orange vertical line followed by a second deep decrease in the MFR to a minimum of about -12\%. A similar feature is also seen in SOPO data, not shown in this figure. Such a two-step feature, however, is not seen at all in data from another polar NM (THUL, blue curve) in which a monotonic decrease starts after the shock arrival, reaching a minimum of about -6\%. In MD data in Figure 1f, monotonic decreases in the sheath region and in the MFR are seen in Nag-V (black curve) and Hob-V (blue curve), but this is unclear for Sao-V (red curve) due to another large amplitude variation superposed. Syo-V (purple curve) shows a narrower decrease during the MFR period. These are all indications that the observed temporal variation of $I_{i,j}(t)$ includes significant contributions from the rigidity dependent anisotropy, which appears different in different directional channels and different rigidity ranges, in addition to the rigidity dependent decrease of GCR density (or ominidirectional intensity). For an accurate analyses of $I_{i,j}(t)$, therefore, it is necessary to analyze the rigidity dependent contributions from density and anisotropy, separately. Such an accurate analysis is possible only with global network data observed by both NMs and MDs.\par

For such analysis considering up to the second order anisotropy representing the bidirectional streaming (BDS), we model $I_{i,j}(t)$ in the geocentric (GEO) coordinate system, as
\begin{equation}
I^{fit}_{i,j}(t) = I^{CG}_{i,j}(t)+\sum_{n=0}^{2}\sum_{m=0}^{n} \{ \xi_c^{n,m}(t) \left( c_{i,j}^{n,m} \cos m\omega t_i - s_{i,j}^{n,m}\sin m\omega t_i \right)+\xi_s^{n,m}(t) \left( s_{i,j}^{n,m} \cos m\omega t_i + c_{i,j}^{n,m}\sin m\omega t_i \right) \},
\end{equation}
where $\xi_c^{0,0}(t)$ is the cosmic-ray density, $\xi_c^{n,m}(t)$ and $\xi_s^{n,m}(t)$ for $1 \leq n \leq 2$, $0 \leq m \leq n$ are the components of comsic-ray anisotropy, $t_{i}$ is the local time in hour at the $i$-th detector, $c_{i,j}^{n,m}$ and $s_{i,j}^{n,m}$ are the coupling coefficients which relate (or ``couple'') the observed intensity in each directional channel with the cosmic-ray density and anisotropy in space and $\omega=\pi/12$. In the GEO coordinate system, we set the $x$-axis to the anti-sunward direction in the equatorial plane, the $z$-axis to the geographical north perpendicular to the equatorial plane and the $y$-axis completing the right-handed coordinate system. $ I^{CG}_{i,j}(t)$ in Eq. (1) is a term representing the contribution to $I^{fit}_{i,j}(t)$ from the solar wind convection and the Compton-Getting anisotropy due to Earth's orbital motion around Sun (see below). We calculate the coupling coefficients $c_{i,j}^{n,m}$ and $s_{i,j}^{n,m}$, as
\begin{equation}
\begin{array}{llllll}
c_{i,j}^{n,m}=\int_{P_c}^{\infty}g_n(p)R(x_i,Z_j,p)P_n^m(\cos(\theta_{i,j}(p)))\cos \{m(\phi_{i,j}(p)-\phi_i)\}dp/\int_{P_c}^{\infty}R(x_i,Z_j,p)dp\\
s_{i,j}^{n,m}=\int_{P_c}^{\infty}g_n(p)R(x_i,Z_j,p)P_n^m(\cos(\theta_{i,j}(p)))\sin \{m(\phi_{i,j}(p)-\phi_i)\}dp/\int_{P_c}^{\infty}R(x_i,Z_j,p)dp
\end{array}
\end{equation}
where $R(x_i,Z_j,p)$ is the response function of the $i$-th detector viewing a zenith angle $Z_i$ at an atmospheric depth $x_i$ to primary cosmic-rays with a rigidity $p$, $P_n^m(\cos(\theta_{i,j}(p)))$ is the semi-normalized spherical function by Schmidt (Chapman and Bartels, 1940), $\phi_i$ is the geographical longitude of the $i$-th detector, $\theta_{i,j}(p)$ and $\phi_{i,j}(p)$ are the geographical asymptotic colatitude and longitude of GCRs with rigidity  $p$ to be detected in the $i,j$ directional channel and $g_n(p)$ is the rigidity spectrum of $\xi_c^{n,m}(t)$ and $\xi_s^{n,m}(t)$. Note that $\xi_s^{n,0}(t)$ in Eq.(1) does not contribute to $I^{fit}_{i,j}(t)$ because $s_{i,j}^{n,0}(t)=0$ in Eq.(2). In this paper, we assume the single power-law spectrum for $g_n(p)$, as
\begin{equation}
\begin{array}{llllll}
g_n(p)&=&(p/p_r)^{\gamma_n}
\end{array}
\end{equation}
where $\gamma_n$ is the power-law index and $p_r$ is the reference rigidity, which we set to be 15 GV and 65 GV as representative rigidities for NMs and MDs, respectively.\par

$ I^{CG}_{i,j}(t)$ in Eq. (1) is also calculated by using the coupling coefficients, setting the anisotropy power-law index to $\gamma_1=0$, as
\begin{equation}
I^{CG}_{i,j}(t)=\xi_x^{CG}(t) \left( c_{i,j}^{1,1} \cos \omega t_i - s_{i,j}^{1,1}\sin \omega t_i \right)+\xi_y^{CG}(t) \left( s_{i,j}^{1,1} \cos \omega t_i + c_{i,j}^{1,1}\sin \omega t_i \right)+\xi_z^{CG}(t) c_{i,j}^{1,0}
\end{equation}
where $\xi_x^{CG}(t)$, $\xi_y^{CG}(t)$ and $\xi_z^{CG}(t)$ are three GEO components of the anisotropy vector 
\begin{equation}
\bm{\xi}^{CG}(t)=(2+\Gamma) \frac{-\bm{V}_{SW}(t)+\bm{v}_E}{c}
\end{equation}
defined with the radial solar wind velocity $\bm{V}_{SW}(t)$ in the OMNI data set\footnote{https://omniweb.gsfc.nasa.gov/ow.html}, the velocity of Earth's revolution around the Sun $\bm{v}_E$ (30 km/s toward the orientation opposite to the y-orientation in the geocentric solar ecliptic (GSE) coordinate system) and the power-law index of GCR energy spectrum $\Gamma=2.7$. Note that $\bm{\xi}^{CG}(t)$ represents a viewing direction with maximum cosmic-ray flux, opposite to the flow direction. $I^{CG}_{i,j}(t)$ in Eq. (4) results in $\sim0.3\%$ diurnal variation of $I_{i,j}(t)$ with a maximum phase depending on the viewing direction. By including this term in Eq. (1), the first order anisotropy terms ($\xi_c^{1,m}(t)$ and $\xi_s^{1,m}(t)$) do not contain contributions from the solar wind convection and the Compton-Getting anisotropy due to Earth's orbital motion.\par

For the first analysis provided in this paper, we simply assume a common $\gamma_n$ for $\xi_c^{n,m}(t)$ and $\xi_s^{n,m}(t)$ with different $m$. This means that the anisotropy phase (the orientation of maximum intensity) is rigidity independent, while the amplitude varies with rigidity. This assumption might be inappropriate, because the first order anisotropy ($\xi_c^{1,m}(t)$ and $\xi_s^{1,m}(t)$), for instance, consists components including parallel and perpendicular diffusion and the diamagnetic drift, all of which might have different rigidity dependences. More rigorous analysis taking account of different spectra for different $m$, however, is out of scope of this first analysis work.\par

We derive the best-fit set of nine parameters ($\xi_c^{0,0}(t)$, $\xi_c^{1,0}(t)$, $\xi_c^{1,1}(t)$, $\xi_s^{1,1}(t)$, $\xi_c^{2,0}(t)$, $\xi_c^{2,1}(t)$, $\xi_s^{2,1}(t)$, $\xi_c^{2,2}(t)$ and $\xi_2^{2,2}(t)$) for each hour of data by solving the following linear equations.
\begin{equation}
\frac{\partial \chi^2}{\partial \xi_c^{n,m}(t)}=\frac{\partial \chi^2}{\partial \xi_s^{n,m}(t)}=0,
\end{equation}
where $\chi^2$ is the residual value of fitting defined as
\begin{equation}
\chi^2=\sum_{i,j}{{\chi_{i,j}}^2}=\sum_{i,j}{\{I_{i,j}(t)-I^{fit}_{i,j}(t)\}^{2}/\sigma_{i,j}^2}
\end{equation}
with $\sigma_{i,j}$ denoting the count rate error of $I_{i,j}(t)$. The best-fit anisotropy in the GEO coordinate system is then transformed to the GSE coordinate system for comparison with the solar wind and IMF data. By changing $\gamma_n$, each between -2.0 and +1.0 in 0.1 steps, we repeat solving the linear Eq. (6) and find $\xi_c^{n,m}(t)$, $\xi_s^{n,m}(t)$ and $\gamma_n$ minimizing $\chi^2$. Thus, the total number of free parameters in our best-fit analysis is 12, i.e., nine $\xi_c^{n,m}(t)$ and $\xi_s^{n,m}(t)$ values at the reference rigidity $p_r$ plus three $\gamma_n$ values for $0 \leq n \leq 2$, and the number of degrees of freedom of the best-fit to 90 $I_{i,j}(t)$ values is 78 (=90-12) when there are no missing $I_{i,j}(t)$ data. In the next section, we present the best-fit results.

\section{Results}\label{sec:result}
\subsection{Best-fit performance}\label{subsec:performance}
Black solid circles in Figure 2a display the observed data by four sample NMs (PSNM, ATHN, THUL and SYOW) in upper panels and by four vertical channels of MDs (Nagoya, Hobart, S\~ao Martinho and Syowa) in lower panels, each as a function of time during the 3 days in Figure 1, while gray curves show fits to data using the 12 best-fit parameters for each hour. Overall, the gray curve tracks the observed data well, indicating that the best-fit parameters successfully reproduce the data. Also shown in Figure 2a by thin black, blue and red curves are the individual contributions to the gray curve from the cosmic-ray density, the first-order anisotropy and the second-order anisotropy, respectively (the gray curve is the sum of three thin curves in each panel). The temporal profile of the density contribution is nearly common (with different amplitudes) for all NMs and MDs, except for some differences for polar NMs (such as THUL and SYOW) with the lowest $P_m$ values, while the profile of the anisotropy contribution is quite different in different detector depending on $P_m$ and the viewing direction of each detector. The purple curve in each panel shows the statistical significance (${\chi_{i,j}}^2=\{I_{i,j}(t)-I^{fit}_{i,j}(t)\}^{2}/\sigma_{i,j}^2$) of the difference between the observed and fit data on the right vertical axis. It is seen that overall the contribution to ${\chi_{i,j}}^2$ is larger for NM data than for MD data.\par

Particularly, the data of PSNM, monitoring a rigidity range in between NMs and MDs, are well reproduced, indicating that the best-fit parameters represent the observed rigidity dependence. The data simultaneously observed by SYOW NM and Syow-V MD are also successfully reproduced giving further support for the reliability of the best-fit analysis. Since the asymptotic viewing directions of SYOW NM and Syow-V MD are similar as described in \ref{subsec:data}, these directional channels can observe roughly the same direction in space with different $P_m$. In other words, the observed difference between $I_{i,j}(t)$ for SYOW NM and Syowa-V MD directly reflects the rigidity dependences of the density and anisotropy. Those rigidity dependences are represented by differences between the best-fit colored curves in two panels of SYOW NM and Syow-V MD.\par

As a quantitative measure of the best-fit performance, we calculate the coefficient of determination with adjusted degrees of freedom, as
\begin{equation}
R^2=1.0-\frac{\frac{1}{N_{dof}-1}\sum_{i,j}{\{I_{i,j}(t)-I^{fit}_{i,j}(t)\}^2}} {\frac{1}{N-1}\sum_{i,j}{\{I_{i,j}(t)-I_{ave}(t)\}^2}}
\end{equation}
where $N$ and $N_{dof}$ are numbers of data available at $t$ and degrees of freedom ($N_{dof}=N-12$), respectively, and $I_{ave}(t)=\sum_{i,j}{I_{i,j}(t)}/N$. $R^2$ shown in the left panel of Figure 2b exceeds 0.9 for nearly the entire FD period.\par

As seen in the right panel of Figure 2b, on the other hand, the minimum reduced $\chi^2$ ($\chi^2$ in Eq.(7) divided by $N_{dof}$) is much larger than one, particularly during the FD period. This results from the actual fluctuation of hourly count rate exceeding the statistical error ($\sigma_{i,j}$ ) which is used to calculate $\chi^2$. One possible source of such fluctuation is local effects such as the snow cover effect and/or instrumental instabilities for some NM data, but there is no reason for those local effects to become larger during the FD period. Another possible source might be the so-called ``cosmic-ray scintillation'' arising from the fluctuation of the magnetic field orientation along which cosmic-rays flow \citep{Owens74}. As will be shown in next subsection, the anisotropy is significantly enhanced during the FD period. According to the enhancement, the intensity fluctuation also increases during the FD period even for the same field fluctuation amplitude. In this case, the minimum reduced $\chi^2$ might be closer to one if we could evaluate the actual fluctuation and add it to the statistical error.\par

The number of MD data used in the best-fit analysis is more than three times larger than the number of NM data, while the count rate error is similar in both NM and MD data. Thus there could be uneven contributions to $\chi^2$ from NMs and MDs. However, significant uneven contributions are not seen between blue and red curves in the right panel of Figure 2b showing the individual contributions to the total $\chi^2$ from NMs and MDs, respectively. This is because the statistical significance is smaller in MD data than in NM data. 

\subsection{Best-fit parameters}\label{subsec:parameters}
Figure 3 shows best-fit parameters obtained from the analysis described in the preceding section. Black solid circles in Figure 3a display the best-fit density ($\xi_c^{0,0}(t)$) on the left vertical axis at 15 GV which is the average $P_m$ monitored by NMs. A clear two-step decrease feature is seen in the density (black curve), i.e. the first $\sim$2\% decrease in the magnetic sheath region between the orange and the left purple vertical lines and the second $\sim$5\% decrease starting a few hours before the MFR period delimited by a pair of vertical purple lines. After the minimum in the first step, the density recovers until the second step decrease starts. Also displayed by blue and red solid circles on the right vertical axis (in the same extent of 7 \% as the left vertical axis) are amplitudes of the first and second order anisotropies ($A_1$ and $A_2$) at 15 GV, respectively, showing strong enhancements of the anisotropy in the MFR period. As seen in Figure 3c, the orientation of the maximum intensity in the second-order anisotropy is clearly aligned to the IMF orientation during the MFR period indicating that this anisotropy is consistent with the BDS. This is also clearly seen in Figure 4 showing the best-fit intensity map due to the second-order anisotropy ($n=2$) in the GSE coordinates. The orientations of maximum intensity indicated by X marks are close to the orientations parallel and anti-parallel to IMF indicated by open and solid circles. In Figure 3a, the maximum $A_2$ (red curve) is as large as $\sim$5 \% and almost comparable to the total decrease of density (black curve) in the second step, indicating that there is only a minor modulation of the intensity for GCRs moving along the IMF in the MFR.\par

In Figure 3a, $A_1$ (blue curve) is also enhanced in the MFR period. As seen in Figures 3b and 4, on the other hand, the orientation of the first-order anisotropy is almost perpendicular to the IMF when the anisotropy is enhanced (see also $n=1$ in Figure 4), possibly indicating the dominant contribution from the $\bf{B} \times \bf{G}$ diamagnetic anisotropy in the IMF $\bf{B}$ and the spatial gradient $\bf{G}$ of GCR density.\par

Figure 3d shows the temporal variation of the obtained power-law index ($\gamma_n$). Although there are large fluctuations seen particularly when the anisotropy amplitude is small, the following systematic trends can be seen. First, $\gamma_0$ (black curve) in the MFR is nearly constant at $\sim$-1.2, while it reduces to -2.0 at the local maximum of density before the MFR period, indicating the softer rigidity spectrum during the density recovery preceding the second decrease. Second, $\gamma_2$ (red curve) also tends to be $\sim$-1.2 around the observed maximum of $A_2$. Third, $\gamma_1$ (blue curve) in the MFR significantly decreases from $\sim+0.5$ to $\sim-1.5$ through $\sim-1.2$ when $A_1$ is maximum. These results will be discussed in the next section.\par

\section{Summary and discussion} \label{sec:discussion}
By analyzing the NM and MD data together, we found the power-law indices ($\gamma_n$) of GCR density and anisotropy dynamically changing during a large FD period in 3-5 November, 2021. This implies that the temporal variations of the GCR density and anisotropy would look very different when we analyze this event by using only NM data or MD data. This is actually seen in Figure 3e showing the best-fit density and anisotropy amplitudes at 65 GV which is the average $P_m$ monitored by MDs. The following significant difference from Figure 3a at 15 GV is evident. While  the enhancement of the second-order anisotropy is seen with smaller amplitude, it is broader than Figure 2a starting before the MFR period possibly suggesting the contribution from the anisotropy which is not the BDS. The enhancement of the first-order anisotropy is more prominent before and after the central MFR period than in Figure 3a. Although detailed discussion about the physical processes responsible for all features in Figure 3 is not the purpose of this paper, the following features can be seen.\par

%We also performed the same best-fit analyses by using NM and GMDN data separately and realized that it is hard to derive systematic variations of $\gamma_n$ like those in Figure 2d due to large fluctuations.\par

The rigidity dependence of the FD has been analyzed by using NM and MD data in many studies and it is well known that the density depression in FD decreases with increasing rigidity \citep{Suda79, Nishida83, Yasuno85, Grigo22}. This is consistent with $\gamma_0$ in Figure 3(d) (black curve) staying between -1.5 and -1.0 when significant depressions of the density ($\xi_c^{0,0}(t)$) are observed. A clear two-step decrease feature is observed in the GCR density at 15 GV, i.e. the first $\sim$2 \% decrease in the turbulent magnetic sheath after the IP-shock arrival and the second $\sim$5 \% decrease in the following MFR. As seen in Figure 3e, the GCR density at 65 GV also shows the two-step decrease, but it is different from that at 15 GV in Figure 3a. The density depression in the first step in the sheath period in Figure 3e recovers to $\sim$0 \% before the MFR period, but it is still $\sim$-1 \% at 15 GV in Figure 3a. This earlier recovery at higher rigidity indicates the softening of rigidity spectrum, which corresponds to $\gamma_0$ (black curve) in Figure 3d decreasing toward the local maximum of density before the MFR period.\par

By analyzing cosmic-ray data observed by 14 NMs during a large number of ICME event periods, \citet{Jordan11} claimed that the traditional model of FDs as having one or two steps should be discarded. As shown in Figure 1, however, the temporal variation of GCR intensity appears quite different in different detectors depending on a detector's viewing direction and rigidity response. Therefore, even for a large event like one analyzed in the present paper, it is rather difficult to clarify whether the density is decreasing in two steps or not, without quantitative analyses such as given in this paper.\par

Another interesting feature of the density is the gradual increase preceding the IP shock arrival, which is seen more clearly in Figure 3e at 65 GV\citep{Belov95, Kadokura86}. During this period, $\gamma_0$ (black curve) and $\gamma_1$ (blue curve) in Figure 3d are between -0.7 and +0.2 and between -0.7 and +0.4, respectively, both around 0.0. This is consistent, at least qualitatively, with the shock reflected GCRs gaining an energy boost ($\Delta E$) through the head-on collision with the shock front. Since the energy boost relative to GCR energy ($E$) is energy independent, the expected excess intensity of GCRs from the shock front ($\Delta I/I = \Gamma \Delta E/E=2\Gamma V_{SW}/c$) and the density given by averaging the excess intensity are both energy independent with $\gamma_1=\gamma_0=0$. Also the GSE longitude of the first order anisotropy during this period in Figure 3b is between $0^\circ$ and $62^\circ$ and consistent with the shock reflection anisotropy from the solar wind upstream direction.\par 

One of the most striking features of this event is an unusually large amplitude BDS observed around the center of the MFR period, indicating a significant population of cosmic-rays moving along the IMF. The obtained power-law index ($\gamma_2$) of the second-order anisotropy is $\sim$-1.2 similar to $\gamma_0$ of the GCR density at around the center of the MFR period. Since this spectrum is much harder than that observed by \citet{Ruffolo06} in October 1989 event, the BDS observed in this event is probably due to GCRs, not solar energetic particles, trapped inside the MFR. There is also no coincident ground level enhancement reported from the world network of NMs, only one on October 28 which was almost a week before the BDS was observed. The maximum intensity excess along the IMF in Figure 3(a) is $\sim+5$ \% at 15 GV relative to the omnidirectional intensity and is comparable to the density decrease in the MFR.\par

The deficit of particles near pitch angle 90$^\circ$ might be expected from the adiabatic focusing in an expanding MFR, which selectively reduces the intensity of GCRs with non-zero perpendicular momentum (${p_\bot}$) keeping the adiabatic constant (${p_\bot}^2/B$) unchanged. The intensity reduction, $\Delta I/I$,   is proportional to the relative loss of momentum, $\Delta p/p$,  which is determined by the product of the rate of deceleration and the time GCRs spent trapped in the expanding and weakening  magnetic field. For betatron deceleration, the rate of cooling is proportional to the momentum, so that $p_\bot/p$ is independent of rigidity. This leads to the suggestion that the $p^{-1}$ spectrum of the intensity reduction is caused by the less effective trapping of the higher rigidity GCRs in accord with the different time profiles seen by NMs and MDs. The result, that the power-law indices of the density depression and the BDS are both $\sim$-1.2 and the maximum amplitude of BDS is comparable to the magnitude of the density depression inside the MFR, can be interpreted most naturally if the intensity of GCRs moving along the IMF is almost free from the modulation, i.e., nearly equilibrated with the GCR flux outside the MFR. As noted above, it is commonly observed that such equilibration in density is faster at higher rigidity, i.e., that $\gamma_0<0$, so it is reasonable that the equilibration of anisotropy is also faster, with $\gamma_1<0$ and $\gamma_2<0$ as well.\par

Selective deceleration of GCRs with non-zero perpendicular momentum is also expected from the adiabatic cooling in an expanding MFR. Let us approximate a local part of the MFR by a straight cylinder along z-axis expanding with velocities, $U_x$, $U_y$ and $U_z$. Then the pitch angle ($\Theta$) dependence of adiabatic deceleration would be proportional to $(\partial U_z/\partial z)\cos^2\Theta+(1/2)(\partial U_x/\partial x+\partial U_y/\partial y)\sin^2\Theta$. Since it is reasonable to assume that the MFR length along z-axis increases linearly with increasing the radial distance ($r$) from Sun, while the lateral dimensions overexpands, increasing faster than $r$, we would expect $\partial U_x/\partial x, \partial U_y/\partial y > \partial U_z/\partial z$ and that the adiabatic cooling reduces the flux near $\Theta=90^\circ$ selectively.\par

%According to the MHD simulation of this event\footnote{see the movie available at https://www.ngdc.noaa.gov/enlil/}, the ICME analyzed in this paper is formed by two ICMEs merged before arriving at Earth. A fast ICME overtook the slower ICME ahead and the coalescence was completed at around the time when the IP-shock is recorded at Earth in late November 3. It might be possible that the magnetic reconnection between two MFRs triggered the injection of GCRs from outside MFRs and shock.\par

In any case, the dominant mechanism that is responsible for the unusually large BDS in this event is not certain. It may be possible that the large amplitude BDS of GCRs exists also in other MFRs, but it can be observed only close to the central axis of the MFR. In this case, the observation of a large amplitude BDS indicates that the ``impact factor'' of Earth to the central axis is very small in this event. If the pitch angle scattering dominates during the GCR propagation, GCRs enter the MFR through the perpendicular diffusion and lose energy due to the MFR expansion. In this case, it would be difficult to maintain the significant populations of GCRs near 0$^\circ$ and 180$^\circ$ pitch angles that are observed in the present event, because such populations will rapidly spread to other directions by the pitch angle scattering, unless there is a particular injection mechanism to supply field-aligned GCRs into the MFR. If the magnetic field is so smooth that the pitch angle scattering is negligible, on the other hand, GCR population near 0$^\circ$ and 180$^\circ$ pitch angles can be maintained when GCRs are supplied from outside and trapped inside for a sufficiently long period. Based on numerical simulations of GCR propagation into the model MFR, \citet{KandR09} reported that GCRs can enter the MFR by the guiding center drift, predominantly along a leg of the MFR in regions of phase space with low ${p_\bot}^2/B$, so the incoming GCRs are concentrated near 0$^\circ$ and 180$^\circ$ pitch angles, and can remain trapped inside the MFR for longer than 25 hours. This might be the case when a large amplitude BDS in the MFR is observed. Analyses of other MFR events are planned to further clarify these interpretations.\par

We finally note a significant softening of the first-order anisotropy in the MFR period as indicated by $\gamma_1$ (blue curve in Figure 3(d)). So far $\gamma_1\sim0$ has been assumed in analyses of the anisotropy observed by GMDN (e.g. \citet{Kihara20}) based on the diffusive GCR transport picture in which the first-order anisotropy is expressed in terms of the diffusion balancing with the rigidity independent solar wind convection. However, $\gamma_1$ in Figure 3(d) might be indicating that the diffusive transport picture is not appropriate in the MFR period of this event. This is probable particularly in the case of the weak pitch angle scattering in the MFR as discussed above. The softening of the first-order anisotropy obtained in this paper is qualitatively consistent with one conclusion of the pioneering work by \citet{Richardson00}. They mentioned that the first-order (or unidirectional) anisotropy in low and high energy ranges can be weakly correlated because they are influenced by the connection to particle sources, which can be different at low and high energies, and by density gradients within particle populations. Further study is also needed to clarify the physical origin of the rigidity dependent first-order anisotropy.\par

\begin{acknowledgments}
This study is a part of the Science Program of Japanese Antarctic Research Expedition (JARE) Prioritized Research Project (Space environmental changes and atmospheric response explored from the polar cap) which was supported by National Institute of Polar Research (NIPR) under MEXT. It is supported in part by the joint research programs of the National Institute of Polar Research (AJ1007), the Institute for Space-Earth Environmental Research (ISEE), Nagoya University, and the Institute for Cosmic Ray Research (ICRR), University of Tokyo in Japan. This work is also partially supported by ``Strategic Research Projects'' grant from ROIS (Research Organization of Information and Systems). The observations are supported by Nagoya University with the Nagoya muon detector, by INPE and UFSM with the S\~{a}o Martinho da Serra muon detector, by the Australian Antarctic Division with the Hobart muon detector, and by project SP01/09 of the Research Administration of Kuwait University with the Kuwait City muon detector. N. J. S. thanks the Brazilian Agency - CNPq for the fellowship under grant number 300886/2016-0. EE would like to thank Brazilian funding agencies for research grants FAPESP (2018/21657-1) and CNPq (PQ-301883/2019-0). M. Rockenbach thanks the Brazilian Agency - CNPq for the fellowship under grant number 306995/2021-2. ADL thanks CNPq for grant 309916/2018-6. Research in Thailand was supported by grant RTA6280002 from Thailand Science Research and Innovation. PSNM maintenance was also supported by Achara Seripienlert and the National Astronomical Research Institute of Thailand. We acknowledge the NMDB database (\url{http://www.nmdb.eu}), founded under the European Union's FP7 programme (contract no. 213007) for providing data. We also gratefully acknowledge the NOAA Air Resources Laboratory (ARL) for the provision of GDAS data, which are available at READY website (\url{http://www.ready. noaa.gov}) and used in this paper. The OMNIWeb dataset of the solar wind and IMF parameters is provided by the Goddard Space Flight Center, NASA, USA.

%while the MHD simulation of the space weather during the event studied in this paper is provided by the WSA-Enlil solar wind prediction, National Centers of Environmental Information, NOAA, USA. 
\end{acknowledgments}

\newpage

\appendix
\section{Response functions used in this paper}
The response function $R(x,Z,p)$ gives the count rate of the atmospheric neutrons or muons produced by primary GCRs with the rigidity $p$ and detected with the incident zenith angle $Z$ at the atmospheric depth $x$. In this paper, we use the $R(x,Z,p)$ by \citet{Nagashima89} for solar minimum conditions for calculating NM characteristics and that by \citet{Murakami79} for MD characteristics in Table 1. We assume $Z=0$ for each NM which is an omnidirectional detector monitoring the vertical incident direction on average. Dashed and solid curves in Figure 5 (a) display $R(x,Z,p)$ of a sample of four NMs and four vertical directional channels of MDs, respectively, each as a function of primary GCR rigidity $p$. It is seen that a wide range of primary rigidities between $\sim1$ GV and  $\sim100$ GV is covered by observations with NMs and MDs. PSNM (red dashed curve) is a unique detector monitoring the rigidity range in between ranges monitored by NMs and MDs. By using this $R(x,Z,p)$, we calculate $P_m$ as $p$ below which the integrated rigidity response is a half of  the total integrated response, as
\begin{equation}
0.5= \int_{P_c}^{P_m}R(x,Z,p)dp/ \int_{P_c}^{\infty}R(x,Z,p)dp.
\end{equation}
As shown in Figure 5 (b), $P_m$ is calculated as $p$ where $R(x,Z,p)$ integrated below $p$ crosses 0.5 on the vertical axis indicated by the horizontal line. The asymptotic viewing directions ($\lambda_{asymp}$ and $\phi_{asymp}$) in Table 1 are then calculated by tracing the orbit of a GCR with $P_m$ in the IGRF-13 model magnetosphere \citep{Lin95}. The same $R(x,Z,p)$ is also used for calculating coupling coefficients in Eqs. (2) and (4).

\newpage

\begin{deluxetable}{ccccccccccc}
\tabletypesize{\scriptsize}
\tablewidth{0pt}
\tablenum{1}
\tablecaption{Characteristics of neutron monitors and muon detectors used in this paper\label{tab:CR data}}
\tablehead{
\colhead{name} & \colhead{$\lambda_D (^{\circ})$} & \colhead{$\phi_D (^{\circ})$} & \colhead{alt. (m)} & \colhead{ch-no.} & \colhead{$P_c$ (GV)} & \colhead{cph/$10^4$} & \colhead{$\sigma$ (0.01\%)} & \colhead{$P_m$ (GV)} & \colhead{$\lambda_{asymp} (^{\circ})$} & \colhead{$\phi_{asymp} (^{\circ})$}
%\colhead{} & \colhead{($^{\circ}$)} & \colhead{($^{\circ}$)} & \colhead{} & \colhead{(m)} & \colhead{(GV)} & \colhead{} &  \colhead{(\%)} & \colhead{(GV)} & \colhead{($^{\circ}$)} & \colhead{($^{\circ}$)}
}
%\decimalcolnumbers
\startdata
&&&&&&21 NMs\\
\tableline
APTY  &   67.6N &   33.4E &  181 & 1 & 0.7 &   68.3 & 12.1 & 15.0 & 41.3N & 64.2E\\
ATHN  &   38.0N &   23.8E &  260 & 1 & 8.5 &   20.7 & 22.0 & 22.8 &   3.7N & 82.0E\\
BKSN  &   43.3N &   42.7E & 1700 & 1 & 5.6 &   42.5 & 15.3 & 16.7 &   6.0S & 103.5E\\
CALM  &   40.6N & 356.8E &  708 & 1 & 7.0 &   25.8 & 19.7 & 20.4 &   3.2N & 57.0E\\
DRBS  &   50.1N &     4.6E &  225 & 1 & 3.2 &   39.4 & 15.9 & 15.5 &   3.5N & 56.0E\\
FSMT  &   60.0N &  248.1E &  203 & 1 & 0.3 & 100.3 & 10.0 & 15.1 & 33.1N & 269.6E\\
INVK   &   68.4N &  226.3E &    21 & 1 & 0.3 &   74.9 & 11.6 & 15.1 & 45.2N & 242.1E\\
IRK2   &    52.4N & 100.6E & 2000 & 1 & 3.6 & 136.1 & 8.6 & 14.0 &   2.8N & 149.6E\\
JNG1  &    46.6N &    8.0E & 3475 & 1 & 4.5 & 115.2 & 9.3 & 13.5 &   9.7S & 69.9E\\
KERG  &    49.4S &  70.3E &     33 & 1 & 1.1 &   83.2 & 11.0 & 14.9 & 10.9S & 83.0E\\
LMKS  &    49.2N &  20.2E & 2634 & 1 & 3.8 &  161.9 & 7.9 & 13.5 &  4.2S & 73.9E\\
MXCO &    19.8N & 260.8E & 2274 & 1 & 8.2 &  84.7 & 10.9 &  20.4 & 11.6S & 327.0E\\
NAIN   &    56.6N & 298.3E &    46 & 1 & 0.3 &  85.1 & 10.8 &  15.1 & 27.3N & 338.8E\\
OULU  &    65.1N &   25.5E &    15 & 1 & 0.8 &  37.1 & 16.4 &  14.9 & 35.2N & 58.4E\\
PWNK  &   55.0N &  274.6E &    53 & 1 & 0.3 &  86.5 & 10.8 &  15.1 & 26.7N & 307.4E\\
SOPO  &   90.0S &        -   & 2820 & 1 & 0.1 &121.4 & 9.1 &  11.3 & 54.7S & 344.5E\\
TERA  &   66.7S & 140.0E  &     32 & 1 & 0.0 &  47.4 & 14.5 &  14.8 & 67.5S & 161.3E\\
THUL  &   76.5N &  291.3E &     26 & 1 & 0.3 &   47.9 & 14.5 & 15.0 & 67.9N & 322.9E\\
TXBY  &   71.6N &  128.8E &      0 & 1 & 0.5 &    39.4 & 15.9 & 14.9 & 47.2N & 162.1E\\
PSNM &   18.6N &    98.5E & 2565 & 1 & 16.7 & 225.0 & 6.7 &  34.6 &  6.0N & 158.7E\\
SYOW &   69.0S &   39.6E &     25 & 1 &  0.4 &   31.8 & 17.7 & 14.9 & 30.4S & 37.8E\\
\tableline
&&&&&&69 MD directional channels\\
\tableline
Nagoya & 35.2N & 137.0E & 77 & 17 & 8.0-12.6 & 17.3-285.6 & 5.9-24.0 & 58.4-106.9 & 64.0N-24.4S & 89.1E-235.0E\\
Hobart & 43.0S & 147.3E & 65 & 13 & 2.5-4.0 & 19.9-149.3 & 8.2-22.4 & 53.1-74.0 & 5.0N-76.6S & 122.4E-237.0E\\
Kuwait & 29.4N & 48.0E & 19 & 13 & 8.9-14.1 & 12.6-252.0 & 4.7-23.3 & 61.2-104.0 & 79.3N-26.1S & 16.8E-136.2E\\
S\~ao Martinho & 29.4S & 306.2E & 488 & 17 & 7.1-14.1 & 4.3-257.1 & 6.2-48.5 & 54.3-98.4 & 33.4N-67.1S & 100.6W-11.6E\\
Syow-MD & 69.0S & 39.6E & 25 & 9 & 2.51-3.55 & 1.0-26.9 &16.0-55.0 & 55.5-72.0 & 6.2S-75.5S & 11.6E-346.0E\\
\enddata
\tablecomments{First 21 rows describe characteristics of 21 neutron monitors, while  the bottom 4 rows present ranges of corresponding parameters for 5 multidirectional muon detectors including the GMDN (Nagoya, Hobart, Kuwait and S\~ao Martinho) and Syowa MD (Syow-MD). The total number of directional channels used in this work is 90 (21 from NMs and 69 from MDs). From left, each column lists the detector name, geographic longitude ($\phi_D$) and latitude ($\lambda_D$), altitude of detector's location, number of directional channels available from the detector, geomagnetic cut-off rigidity ($P_c$) for each directional channel, average hourly count rate, count rate error ($\sigma$), median rigidity of primary GCRs ($P_m$), geographic longitude ($\phi_{\rm asymp}$) and latitude ($\lambda_{\rm asymp}$) of the asymptotic viewing direction outside the magnetosphere. $P_m$ is calculated by using the response function of each detector to primary GCRs, while $\lambda_{\rm asymp}$ and $\phi_{\rm asymp}$ are calculated by tracing orbits of GCRs with $P_m$ in the model magnetosphere (see Appendix).}
\end{deluxetable}

\newpage

\begin{figure}[ht!]
\plotone{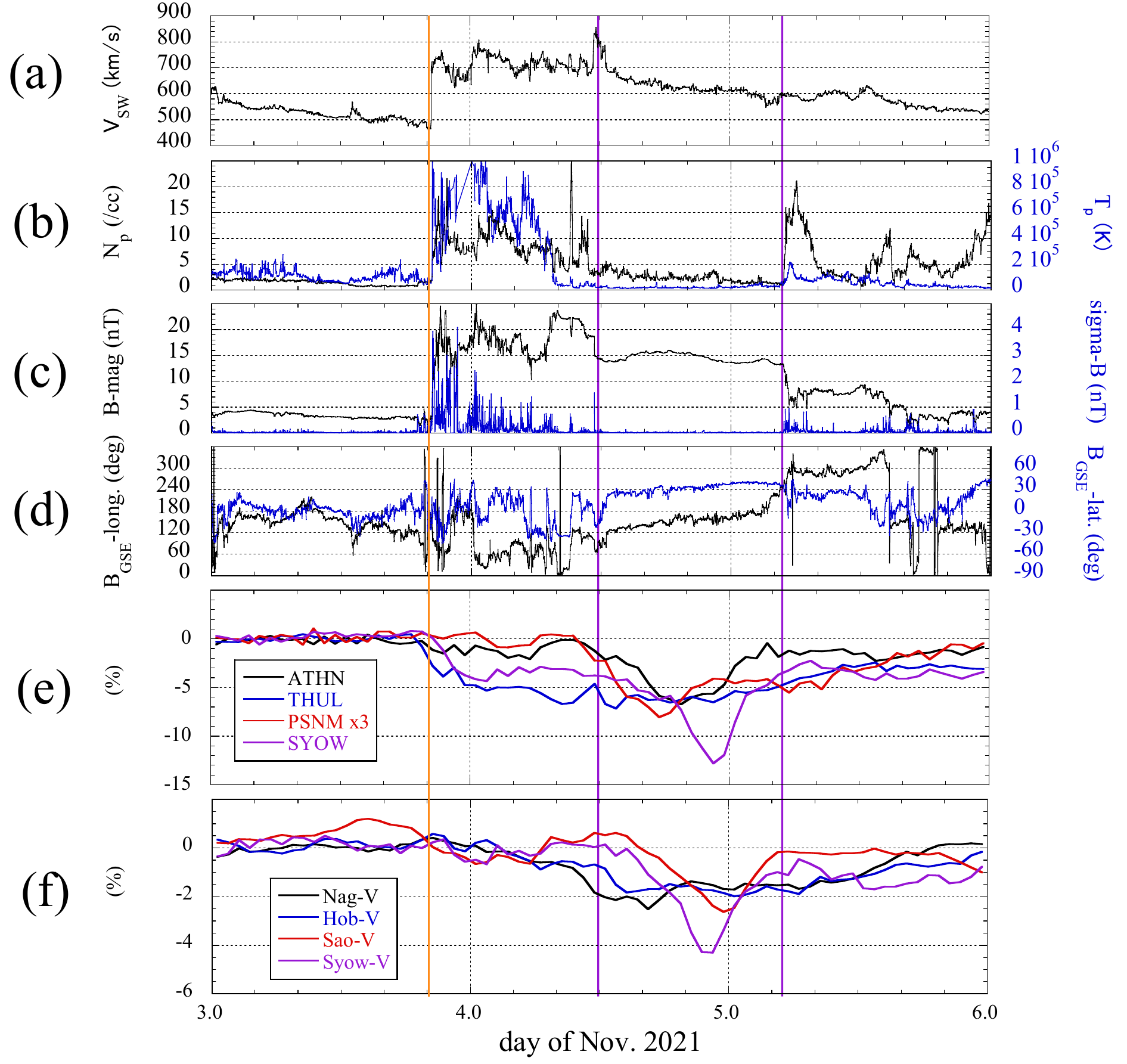}
\caption{Solar wind parameters and cosmic-ray data during 3-5 November, 2021. Panels (a)-(d) show one minute solar wind data; (a) solar wind velocity, (b) plasma density and temperature on the left and right vertical axes, respectively, (c) IMF magnitude and its fluctuation, on the left and right vertical axes, respectively and (d) GSE-longitude and latitude of IMF orientation, on the left and right vertical axes, respectively. Panels (e) and (f) show hourly count rates of a sample of four NMs (ATHN, THUL, PSNM and SYOW) and four vertical channels of MDs (Nagoya, Hobart, S\~ao Martinho and Syowa). PSNM data in (e) are multiplied by three to show the variation more clearly. The vertical orange line indicates the IP-shock arrival, while a pair of purple vertical lines delimit the MFR period defined in this paper.\label{fig:IMF}}
\end{figure}

\newpage

%\begin{figure}[ht!]
\begin{sidewaysfigure}
\plotone{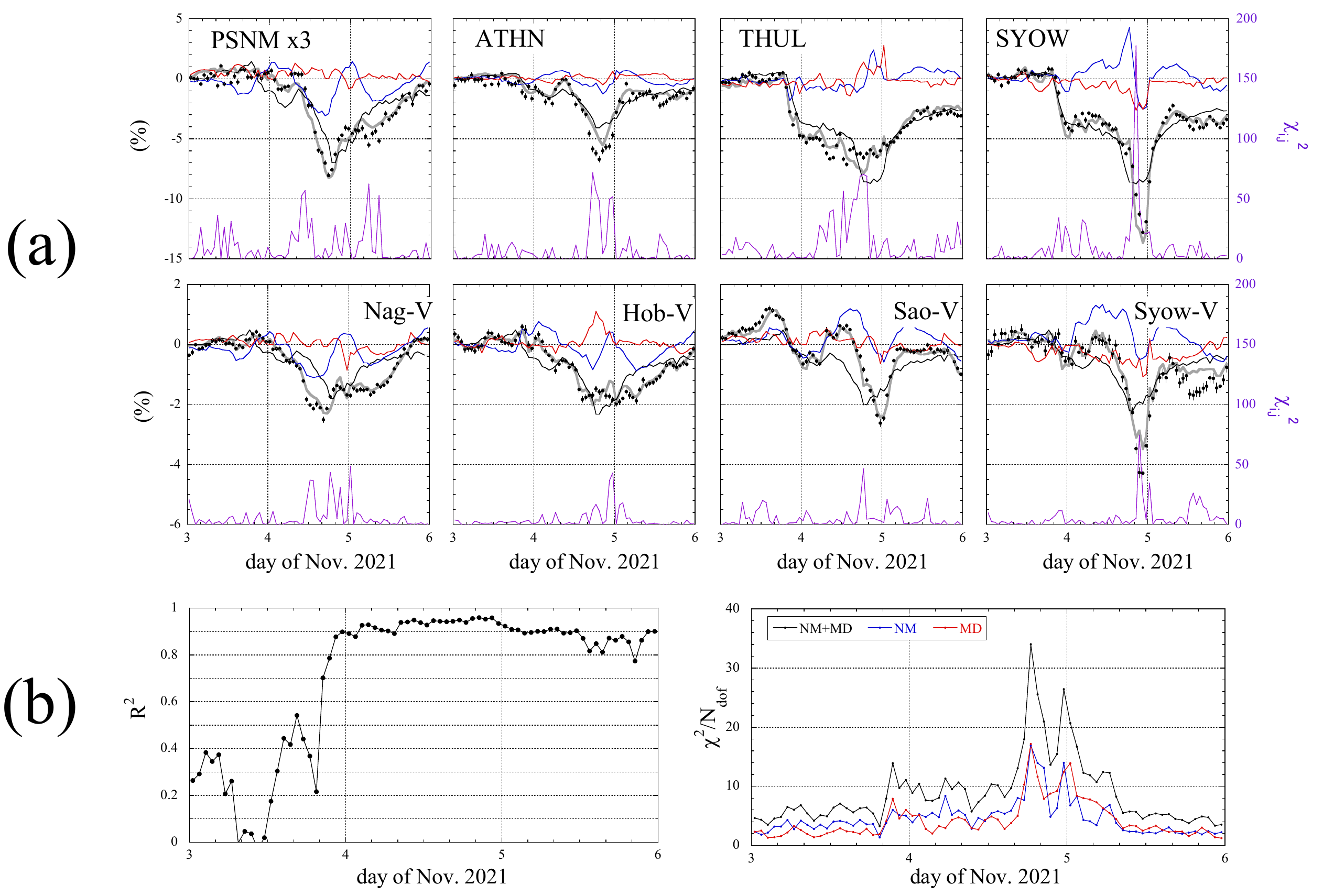}
\caption{Best-fit performance. In each panel of Figure 2(a), black solid circles with statistical error bars and gray curves show the observed and the fit to cosmic-ray data, respectively, for a sample of four NMs (ATHN, THUL, PSNM and SYOW) in upper panels and four vertical channels of MDs (Nagoya, Hobart, S\~ao Martinho and Syowa) in lower panels. PSNM data in the left panel of (a) are multiplied by three. Also shown in each panel are the individual contributions from the GCR density (black thin curve), the first order anisotropy (blue thin curve) and the second-order anisotropy (red thin curve). The gray curve is the sum of three thin curves in each panel. The purple curve at the bottom of each panel is the contribution to $\chi^2$ from that channel. In Figure 2(b), the coefficient of determination $R^2$ (left panel) and the reduced $\chi^2$ (right panel) are shown as functions of time. The reduced $\chi^2$ is calculated by dividing $\chi^2$ in Eq. (7) by the number of degrees of freedom $N_{dof}$ (see Section \ref{subsec:analyses}).\label{fig:repro}}
\end{sidewaysfigure}
%\end{figure}

\newpage

\begin{figure}[ht!]
\plotone{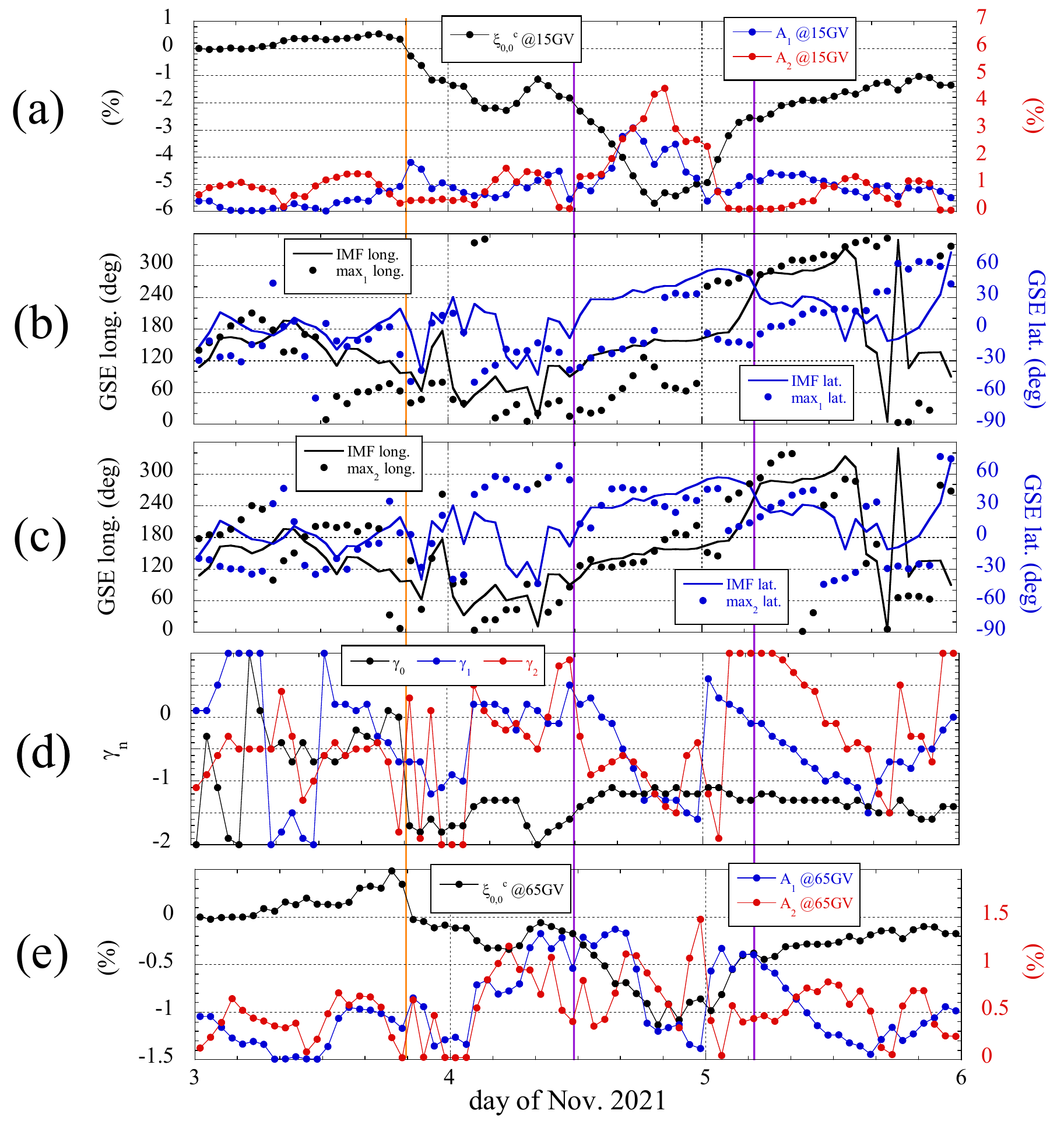}
\caption{Best-fit cosmic-ray parameters in 3-5 November, 2021. Panel (a) shows GCR density (black circles) on the left vertical axis and amplitudes of the first-order (blue circles) and second-order (red circles) anisotropies on the right vertical axes, each at 15 GV which is the representative $P_m$ for NMs. Note an identical extent (7 \%) of the left and right vertical axes. We calculate these amplitudes ($A_1$ and $A_2$) as $A_1=\sqrt { \sum_{m=0}^{1} \{ {\xi_c^{1,m}(t)}^2+{\xi_s^{1,m}(t)}^2 \} }$ and $A_2=\sqrt { \sum_{m=0}^{2} \{ {\xi_c^{2,m}(t)}^2+{\xi_s^{2,m}(t)}^2 \} }$. In panel (b), black (blue) solid circles display the GSE-longitude (latitude) of the orientation of the maximum intensity in the first-order anisotropy on the left (right) vertical axis. Panel (c) also displays the GSE-longitude and latitude of the second-order anisotropy in a same manner as (b). In panels (b) and (c), black and blue solid curves display the longitude and latitude of the IMF orientation, respectively. For the second-order anisotropy, there are two directions of maximum intensity; we plot the direction that is closer to the IMF orientation. Panel (d) shows the power-law indices of rigidity spectra of the density (black), first-order anisotropy (blue) and second-order anisotropy (red).  Panel (e) shows GCR density and anisotropy amplitudes at 65 GV, which is the representative $P_m$ for MDs, in the same manner as (a).\label{fig:bf_param}}
\end{figure}

\newpage

\begin{figure}[ht!]
%\epsscale{.5}
\plotone{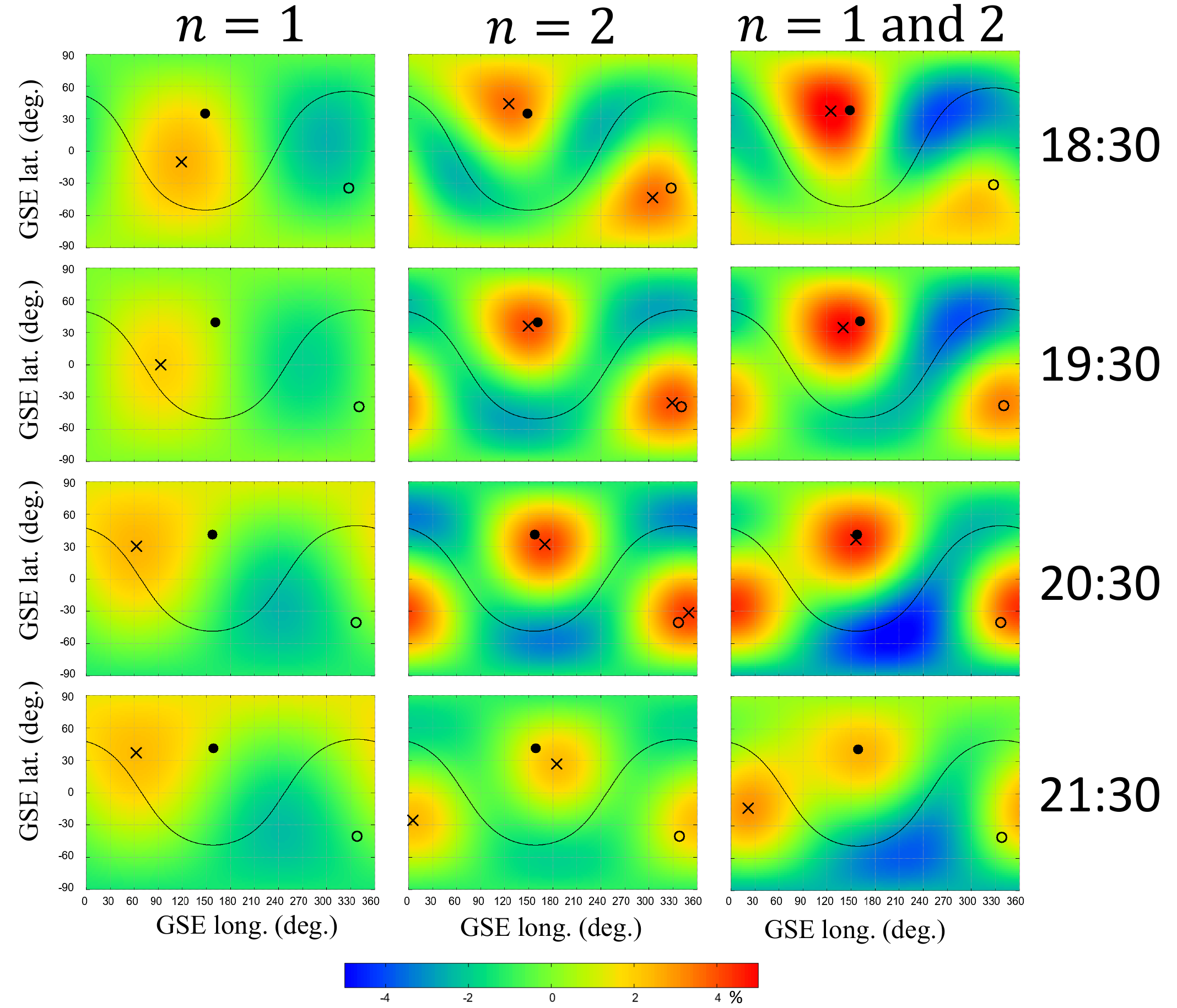}
\caption{Intensity map of the best-fit anisotropy at 15 GV in GSE-coordinates. From the left each column shows the intensity in the first-order ($n=1$), second-order ($n=2$) and total ($n=1$ and $n=2$) anisotropies, each as a color map in the GSE longitude and latitude space, while from the top each raw shows the anisotropy in four hours between 18:30 and 21:30 of Nov. 4 when the large amplitude anisotropy is observed. The solid and open circles indicate the orientations parallel and anti-parallel to the IMF, respectively, while the solid curve indicates the magnetic equator. The X mark indicates the orientation of maximum intensity.\label{fig:GSE_map}}
\end{figure}

\newpage

\begin{figure}[ht!]
\epsscale{.8}
\plotone{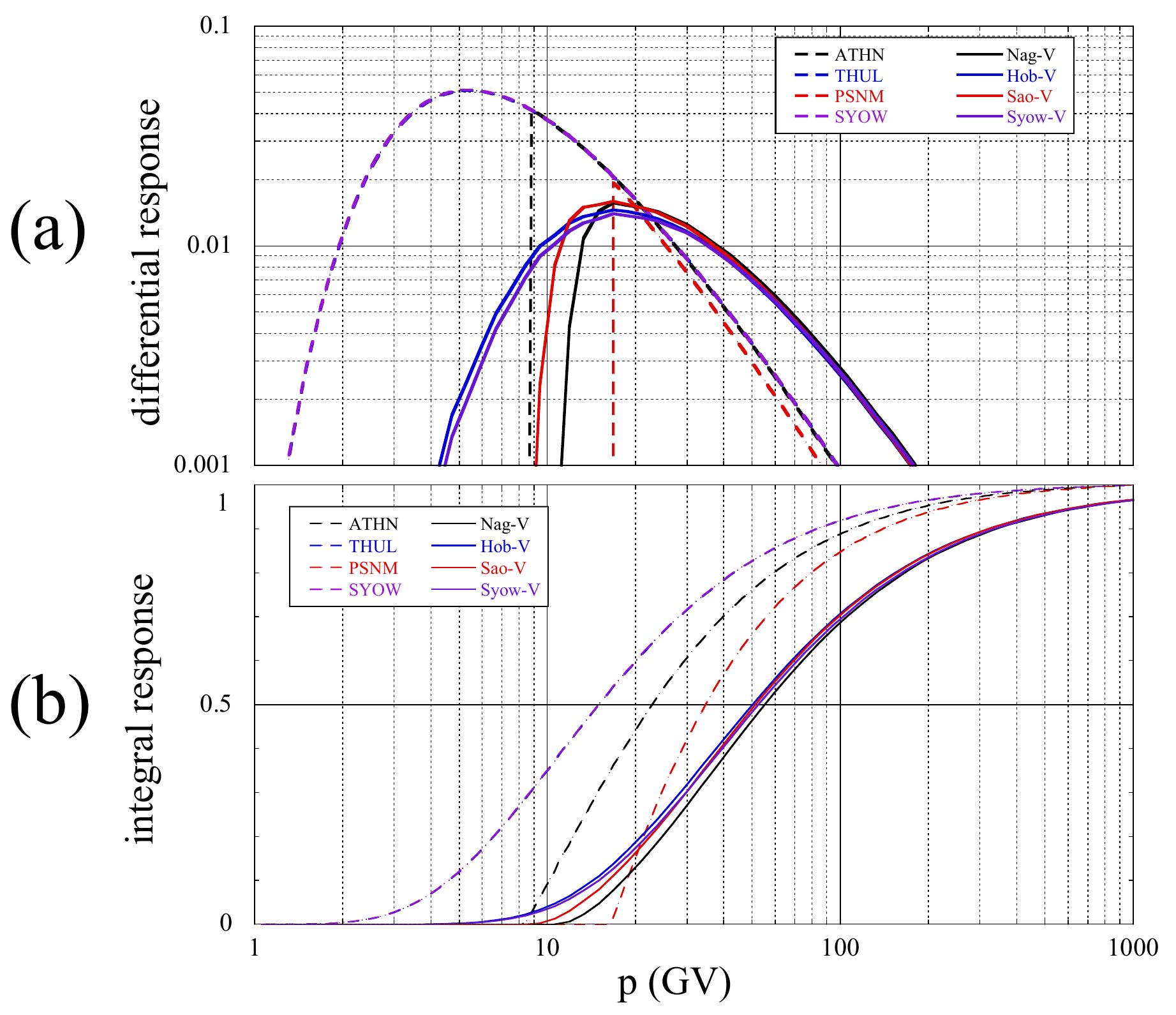}
\caption{Response functions of a sample of four NMs and four vertical channels of MDs. Panel (a) shows the differential response functions $R(x,Z,p)$, while panel (b) displays $R(x,Z,p)$ integrated below $p$, each as a function of primary GCR's rigidity $p$. Black, blue, red and purple dashed curves show response functions of ATHN, THUL, PSNM and SYOW, respectively, while black, blue, red and purple solid curves show functions of four vertical channels of Nagoya, Hobart, S\~ao Martinho and Syowa MDs, respectively. Note that blue and purple curves of THUL and SYOW are overlapped due to similar $P_c$ and atmospheric depth at these NMs. All response functions are normalized and divided by the total integral of $R(x,Z,p)$ over $p$ ($\int_{P_c}^{\infty}R(x,Z,p)dp$).\label{fig:resp}}
\end{figure}

\newpage



\end{document}